\documentclass[preprint]{aastex}

\def\puncspace{\ifmmode\,\else{\ifcat.\C{\if.\C\else\if,\C\else\if?\C\else%
\if:\C\else\if;\C\else\if-\C\else\if)\C\else\if/\C\else\if]\C\else\if'\C%
\else\space\fi\fi\fi\fi\fi\fi\fi\fi\fi\fi}%
\else\if\empty\C\else\if\space\C\else\space\fi\fi\fi}\fi}
\def\SP{\let\\=\empty\futurelet\C\puncspace }
\def\etal{et\SP al.\SP }

\def\h-1{$h^{-1}$}

\def\void#1{{}}

\def\h1{$h^{-1}$}

\def\iras{$IRAS$\SP}
\def\kms{kms$^{-1}$\SP}
\def\etal{et al.\,}
\def\eg{e.g., \,}

\def\lsim{~\rlap{$<$}{\lower 1.0ex\hbox{$\sim$}}}
\def\gsim{~\rlap{$>$}{\lower 1.0ex\hbox{$\sim$}}}
\def\hmu2 {$\bar{\mu}_{1/2}$\SP}

\def\dnsig{$D_n-\sigma$\SP}

\def\rhat{{\bf \hat r}}

\input psfig

\begin{document}

\title{Redshift-Distance Survey of Early-Type Galaxies: Dipole of the
Velocity Field}

\author{L. N. da Costa \altaffilmark{1,}\altaffilmark{2},
M. Bernardi\altaffilmark{1,}\altaffilmark{3,}\altaffilmark{4},
M. V. Alonso\altaffilmark{5}, G. Wegner\altaffilmark{6}}

\author{C. N. A. Willmer\altaffilmark{2,}\altaffilmark{7},
P. S. Pellegrini\altaffilmark{2}, M. A. G. Maia\altaffilmark{2}, 
S. Zaroubi\altaffilmark{4}}

\affil{\altaffiltext{1}{European Southern Observatory,
Karl-Schwarzschild Strasse 2, D-85748 Garching,
Germany}}

\affil{ \altaffiltext{2} {Observat\'orio Nacional, Rua General Jos\'e Cristino
77, Rio de Janeiro, R. J., 20921, Brazil}}

\affil{\altaffiltext{3}{Universit\"{a}ts-Sternwarte M\"{u}nchen,
Scheinerstr. 1, D-81679, M\"{u}nchen, Germany}}

\affil{\altaffiltext{4}{Max Planck Institut f\"ur Astrophysik, 
Karl-Schwarzschild Strasse 1, D-85740, Garching, Germany}}

\affil{\altaffiltext{5}{Observatorio Astr\'onomico de
C\'ordoba,  Laprida  854, C\'ordoba, 5000, Argentina}}

\affil{\altaffiltext{6}{Department of Physics \& Astronomy, Dartmouth
College, Hanover, NH  03755-3528, USA}}

\affil{ \altaffiltext{7}{UCO/Lick Observatory, University of California,
1156 High Street, Santa Cruz,  CA 95064, USA}}

%\today

\begin{abstract}

We use the recently completed redshift-distance survey of nearby
early-type galaxies (ENEAR) to measure the dipole component of the
peculiar velocity field to a depth of $cz \sim 6000$~\kms. The sample
consists of 1145 galaxies brighter than $m_B=14.5$ and $cz
\leq7000$~\kms, uniformly distributed over the whole sky, and 129
fainter cluster galaxies within the same volume. Most of the \dnsig
distances were obtained from new spectroscopic and photometric
observations conducted by this project, ensuring the homogeneity of
the data over the whole sky.  These 1274 galaxies are objectively
assigned to 696 objects -- 282 groups/clusters and 414 isolated
galaxies. We find that within a volume of radius $\sim 6000$~\kms, the
best-fitting bulk flow has an amplitude of $\vert {\bf v_b} \vert =220
\pm 42$~\kms in the CMB restframe, pointing towards
$l=304^\circ \pm 16^\circ $, $b=25^\circ\pm11^\circ$. The error in the
amplitude includes statistical, sampling and possible systematic
errors. This solution is in excellent agreement with that obtained by
the SFI Tully-Fisher survey.  Our results suggest that most of the
motion of the Local Group is due to fluctuations within $6000$~\kms,
in contrast to recent claims of large amplitude bulk motions on larger
scales.

\end {abstract}
\keywords{Cosmology: large-scale structure of universe -- cosmology:
observations -- galaxies: distances and redshifts}

\section{Introduction}

Within the gravitational instability framework for the growth of
cosmic structures, the peculiar velocity field of galaxies and
clusters is a direct probe of density fluctuations of the underlying
mass distribution.  Among several possible statistics that can be
used, measurements of the bulk motion amplitude on different scales
are the simplest and provide, at least in principle, constraints on
the power-spectrum of mass fluctuations. This has motivated several
attempts to measure the dipole component of the local peculiar
velocity field and to determine the volume within which the streaming
motion vanishes in the restframe defined by the Cosmic Microwave
Background radiation (CMB). At this distance the distribution of
matter within the encompassing volume should explain the $\sim
600$~\kms motion of the Local Group relative to the CMB restframe.

Observational searches of large-scale flows date back to the
pioneering work of Rubin \etal (1976). Since then, redshift-distance
surveys have greatly expanded, the data quality has improved
significantly, and several recent attempts have been made using
different techniques and samples (\eg Strauss \& Willick
1995). Despite these efforts, the results remain to a large extent
controversial. The original claim that the flowfield out to
$cz\sim4000$\kms is characterized by a coherent, large-amplitude
$\sim500$~\kms streaming motion (Dressler \etal 1987) relative to the
CMB was revised to incorporate a large concentration of mass, the
so-called Great Attractor (hereafter GA), near $l=310^\circ$,
$b=10^\circ$ (Lynden-Bell \etal 1988).  More recent claims for the
existence of a large amplitude flow $\sim 600$~\kms, with a coherence
length of $\sim$ 100\h1 Mpc (\eg Willick 1990; Mathewson, Ford \&
Buchhorn 1992), suggesting excess power on very large scales, have
also received reconsideration from the following standpoints. First, a
careful re-analysis of the available data yielded a significantly
smaller bulk velocity (Courteau \etal 1993). Second, the analysis of
the independent SFI ($I$-band field spiral) TF-survey led to a
different characterization of the flowfield.  Indeed, the SFI velocity
field shows that the flow is not as coherent as originally envisioned,
exhibiting along the Supergalactic Plane a bifurcation towards the
Great Attractor and Perseus-Pisces, similar to that predicted from
reconstructions of the \iras surveys (\eg da Costa \etal 1996).
Furthermore, the flow within 6000~\kms is characterized by a strong
shear across the volume, in contrast to the picture of a coherent
motion of all structures.  Recent analyses based on the re-calibrated
Mark~III catalogs lead to a roughly consistent picture to that
obtained with SFI (da Costa \etal 1996; Dekel \etal 1999), even though
some discrepancies still remain. For instance, Mark~III yields a
systematically larger amplitude of the bulk motion
$\sim370\pm110$~\kms on scales $\sim$ 5000~\kms as compared to values
$\lsim300$~\kms obtained by applying different techniques to the SFI
sample (da Costa \etal 1996; Giovanelli \etal 1998a). In particular, a
direct fit to the SFI radial velocities yields a bulk velocity of $200
\pm 65$~\kms within the sphere of radius $\sim 6500$~\kms consistent
with that obtained from the SCI cluster sample (Giovanelli \etal
1998b). These results suggest that a significant fraction of the LG
motion is generated on scales $\lsim 6000$~\kms.  While recent direct
measurements of the bulk velocity on larger scales (Dale \etal 1999)
suggest that this may indeed be the case, other works (Lauer \&
Postman 1994; Willick 1999; Hudson \etal 1999) argue for the existence
of large amplitude ($\gsim 600$~\kms) streaming motions out to a depth
of 15,000~\kms.

In this paper we use the recently completed all-sky, homogeneous
redshift-distance survey of early-type galaxies (ENEAR, da Costa \etal
2000, hereafter Paper~I) to study the dipole component of the peculiar
velocity field within $cz\lsim6000$~\kms. Our main goal is to compare
our results using an entirely independent sample to those obtained by
existing Tully-Fisher surveys.

\section {The Sample}
\label{sample}

In the present analysis, we use the ENEAR redshift-distance survey
described in greater detail in Paper~I of this series. Briefly, the
ENEAR sample consists of $\sim$1600 early-type galaxies brighter than
$m_B=14.5$ and with $cz \leq$ 7000~\kms, with \dnsig distances
available for 1359 galaxies. Of these 1145 were deemed suitable for
peculiar velocity analysis (Paper~I). To the magnitude-limited sample
we added 285 galaxies fainter and/or with redshifts $> 7000$~\kms, 129
within the same volume as the magnitude-limited sample. In total, the
cluster sample consists of 569 galaxies in 28 clusters, which are used
to derive the distance relation.  Over 80\% of the galaxies in the
magnitude-limited sample and roughly 60\% of the cluster galaxies have
new spectroscopic and R-band photometric data obtained as part of this
program. Furthermore, repeated observations of several galaxies in the
sample provide overlaps between observations conducted with different
telescope/instrument configurations and with data available from other
authors. These overlaps are used to tie all measurements into a common
system, thereby ensuring the homogeneity of the entire dataset. In
contrast to other samples new observations conducted by the same group
are available over the entire sky.  The comparison between the sample
of galaxies with distances and the parent catalog also shows that the
sampling across the sky is uniform.

Individual galaxy distances were estimated from a direct \dnsig
template relation derived by combining all the available cluster data,
corrected for incompleteness and associated diameter-bias (Lynden-Bell
\etal 1988). The construction of the template relation was carried out
following Giovanelli \etal (1997). From the observed scatter of the
template relation the estimated fractional error in the inferred
distance of a galaxy is $\Delta \sim 0.19$, nearly independent of the
velocity dispersion.

Since early-type galaxies are found preferentially in high-density
regions, galaxies have been assigned to groups/clusters using
well-defined criteria imposed on their projected separation and
velocity difference relative to the center of groups and clusters, as
described in paper~I.  Early-type galaxies in a group/cluster are
replaced by a single object having: (1) the redshift given by the
group's mean redshift, which is determined considering all
morphologies; (2) the distance given by the error-weighted mean of the
inferred distances, for groups with two or more early-types; and (3)
the fractional distance error given by $\Delta / \sqrt(N)$, where $N$
is the number of early-types in the group. In some cases groups were
identified with Abell/ACO clusters within the same volume as the ENEAR
sample and fainter cluster galaxies were added, as described in
Paper~I.

\begin{figure}
\centering
\mbox{\psfig{figure=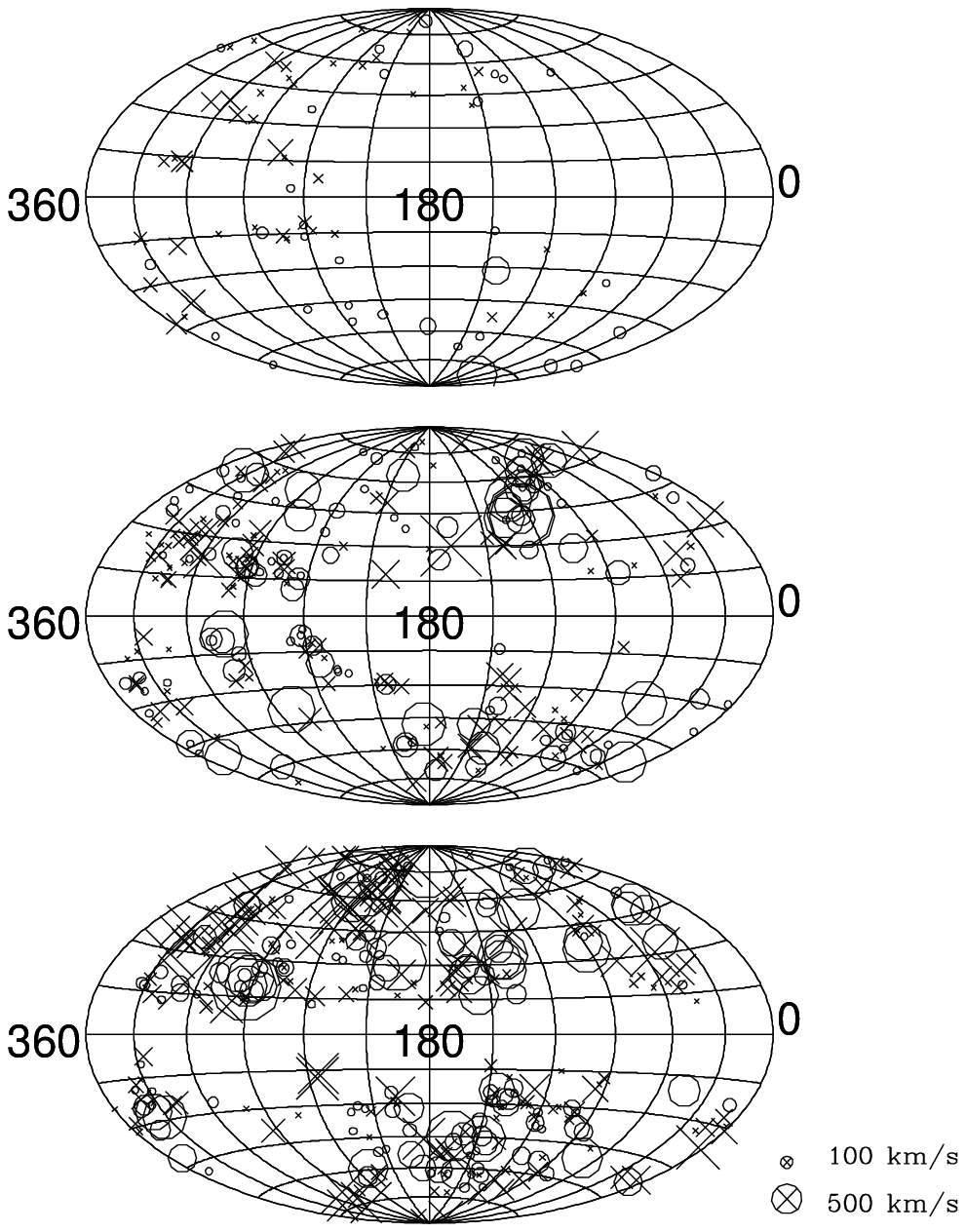,height=15truecm,bbllx=-1.9truecm,bblly=9truecm,bburx=19truecm,bbury=24truecm}}
\figcaption[]{Projected distribution in Galactic coordinates of the
ENEAR peculiar velocity field in different distance shells
2000~kms$^{-1}$ thick in the interval $0 < R < 6000$~kms$^{-1}$. The
velocities are relative to the CMB resframe and the different symbols
represent infall (open circles) and outflow (crosses). Their sizes are
proportional to the galaxy's peculiar velocity
amplitude.\label{fig:all}}
\end{figure}

The inferred distances are corrected for the homogeneous and
inhomogeneous Malmquist bias (IMB). The latter was estimated using the
PCSz density field (Branchini \etal 1999), corrected for peculiar
velocity effects, following Willick \etal (1997). In this calculation
we also include the correction for the sample redshift limit. It
should be noted that this is an approximation as early-types are
biased relative to \iras galaxies. A complete description of the
sample used and the corrections applied will be presented in a
subsequent paper of this series.  As an illustration of the velocity
field mapped by the ENEAR objects we show in Figure~\ref{fig:all} the
projected distribution of objects in Galactic coordinates with the
sample split into different distance shells.  The different symbols
distinguish between objects with positive (crosses) and negative
peculiar velocities (circles). The peculiar velocities are relative to
the CMB restframe and have been computed from fully corrected
distances as described above. For an alternative view of the data we
refer the reader to paper~I. In Figure~\ref{fig:all} structures such
as the GA at $l\sim300^\circ$, $b\sim30^\circ$ and the Perseus-Pisces
(PP) complex at $l\sim120^\circ$, $b\sim-40^\circ$ are easily
recognized in the two outermost shells. Note that in these directions
one finds evidence of outflow and infall as expected around mass
concentrations.  As will be shown in a later paper of this series, the
presence of mass concentration in the PP region is confirmed from the
reconstruction of the three-dimensional velocity field and mass
distribution which shows that both structures have comparable peak
density contrasts. The prominence of the PP complex is perhaps the
most significant difference between the reconstructions based on the
ENEAR and the 7S samples. The ENEAR reconstructed fields are also in
good agreement with those obtained from the PSCz redshift survey
(Branchini \etal 1999), corrected for peculiar velocities, as it will
be shown in a forthcoming paper.

\section {Measurements of the Bulk Motion}
\label{motion}

One of the primary goals of the ENEAR survey has been to investigate
the robustness of previous peculiar velocity analyses using an
independent and uniform sample of early-type galaxies probing a
comparable volume as the recently completed TF-surveys. While many
tests are possible and will be explored in more detail in separate
papers (\eg Borgani \etal 2000), here we consider the dipole component
of the velocity field. A bulk flow model is the simplest way to
globally characterize the velocity field, having been extensively used
in previous work (\eg Dekel 1999 for a recent review). To determine
the best-fitting bulk flow we minimize (\eg Lynden-Bell \etal 1988)
\begin{equation} 
\chi^2 =  \sum w_i \left( u_i -{\bf v}_b \cdot \rhat_i
\right)^2  \label{eq:bulk}
\end{equation} 
where $u_i$ is the radial component of the peculiar velocity of the
$i^{th}$ object in the CMB restframe, located in the direction
$\rhat_i$, ${\bf v}_b$ is the bulk flow and $w_i$ is the weight given
to the $i^{th}$ object in the sample. In our calculations we use
either uniform (equal) weights $w_i=1$ or $w_i =
\frac{1}{\epsilon_i^2 + \sigma^2}$, where $\epsilon_i$, is the sum in
quadrature of the distance and redshift errors (the latter is
negligible in the case of field objects), and $\sigma$ is the
one-dimensional velocity dispersion due to true velocity noise
generated on small scales.

Table~\ref{tab:bulk} summarizes the bulk flow results obtained using
various sub-samples extracted from the combined sample of 696 objects
within different volumes.  The table gives for each volume of radius
$R$ in units of \kms, the number of objects in each sub-sample, the
amplitude and direction of the best-fitting bulk motion, and their
respective errors, obtained using different weighting schemes. The
amplitude of the bulk motion is relative to the CMB restframe and its
direction is expressed in terms of the galactic longitude and
latitude.  The errors were estimated from 1000 Monte-Carlo
realizations generated by adding Gaussian random deviates of the
distance errors to the original distances, from which the dispersion
of the dipole components are calculated. In the table, the weighted
solutions assume a thermal component of $\sigma_f=250$~\kms which is
combined with the object's distance errors in quadrature. The bulk
amplitudes listed in Table~\ref{tab:bulk} have been corrected for the
error-bias obtained subtracting from the square of the best-fitting
value of the bulk velocity the sum in quadrature of the errors in each
Cartesian component (Lauer \& Postman 1994). The amplitude of this
correction is relatively small $\sim$ 50 \kms. We point out that the
amplitude of the bulk velocity at 6000~\kms is insensitive to the
Malmquist bias correction. The comparison between the results obtained
using raw distances, with those corrected only for homogeneous
Malmquist bias (HMB) and those obtained using the full correction, are
comparable to the estimated errors in the bulk velocity. Typical
values for the HMB and IMB corrections are 13\% and 4\%,
respectively. Only the direction of the dipole shows some dependence
on the adopted correction. In particular, neglecting the IMB
correction yields lower values of $b$.  The good agreement between the
direction of the fully corrected ENEAR and those of SFI and Mark~III,
using different procedures to estimate the IMB, is reassuring.

\begin{figure}
\centering
\mbox{\psfig{figure=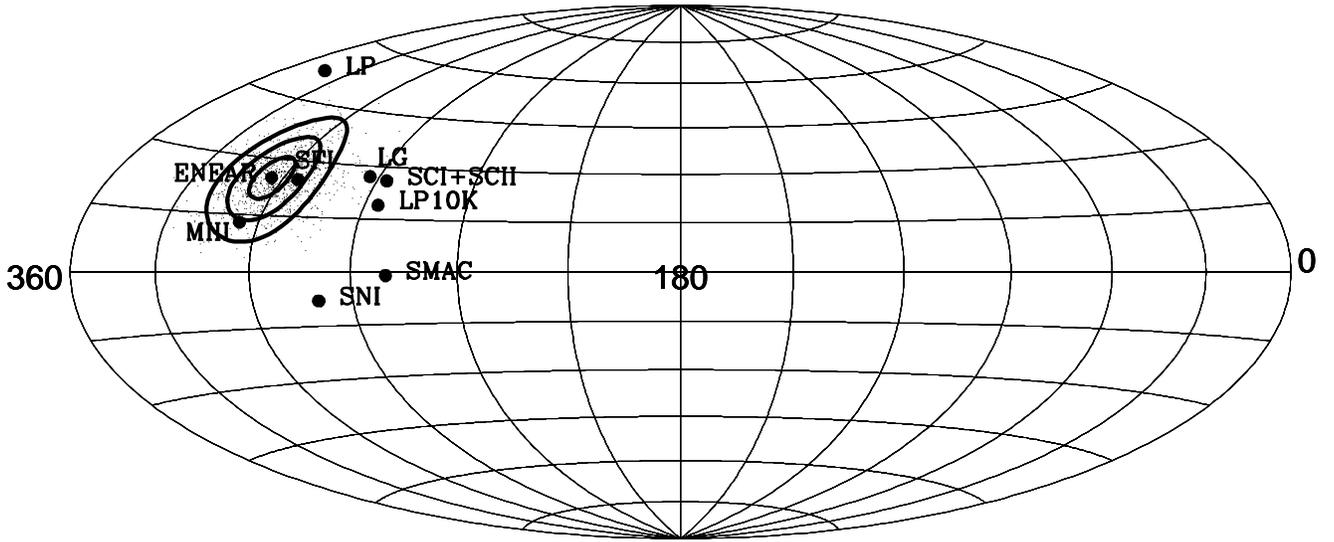,height=11truecm,bbllx=3truecm,bblly=14truecm,bburx=19truecm,bbury=25.5truecm}}
\figcaption[]{The bulk flow direction in Galactic coordinates and the
direction obtained from 1000 Monte-Carlo realizations (dots). The
contours represent 1, 2, and 3$\sigma$ error ellipsoids as derived
from the Monte-Carlo realizations. The figure shows the direction of
the LG motion (LG) and the dipole directions obtained by other authors
on different scales (see text). We adopt the following notation: LP
(Lauer \& Postman 1994); MIII (Dekel \etal 1999); SFI (Giovanelli
\etal 1998a); LP10K (Willick 1999); SCI+SCII (Dale \etal 1999); SNI
(Riess \etal 1997); SMAC (Hudson \etal 1999).\label{fig:simul}}
 
\end{figure}

From the direct fit of the radial velocities using equal weights we
find $|{\bf v}_b|=220 \pm 42$ in the direction $l=304^\circ \pm
16^\circ$, $b=25^\circ\pm 11^\circ$ within a radius of $cz \sim 6000$
\kms. Note that this value is smaller than the preliminary value
reported earlier by Wegner \etal (1999) which was not corrected for
the error-bias and was determined before the full sample had been
assembled. A somewhat larger value is obtained when objects are
weighted by their distance errors, but the amplitude is still less
than 300~\kms and essentially in the same direction. The direction of
the ENEAR dipole is compared in Figure~\ref{fig:simul} to other recent
estimates measured on similar scales ($\sim5000-6500$~\kms) using the
SFI (Giovanelli \etal 1998a) and the revised Mark~III (Dekel \etal
1999) samples. The contours represent the 1-3$\sigma$ confidence
levels, derived from the Monte-Carlo simulations. Perhaps the most
interesting result is the excellent agreement both in direction and
amplitude between the ENEAR and SFI dipole solutions, probably the two
most homogeneous all-sky samples currently available for the analysis
of peculiar velocity data. Particularly important is the well known
fact that early-type (E and S0) and late-type (Sc) galaxies probe
distinct regions of the galaxy distribution - while spirals are found
predominantly in low-density regions and are more uniformly
distributed, the distribution of ellipticals is clumpier, delineating
more clearly the most prominent nearby structures. Equally important
is the fact that the peculiar velocities used in the two studies are
based on independent distance relations involving different
measurements and corrections. In Figure~\ref{fig:simul} we also show
the direction of the dipoles recently measured on larger scales. The
results obtained on scales of $\sim 6000$~\kms are consistent, both in
direction and amplitude, with those measured on much larger scales
using the SCI+SCII sample. Combined these results suggest that while
most of the LG motion stems from fluctuations within 6000~\kms some
contribution also comes from larger scales where a better agreement
between the dipole direction and the LG motion is found. It is
important to note, however, that currently there is very little
agreement among various determinations of the dipole on scales $\gsim$
10,000~\kms.

To evaluate the possible impact of sampling effects directly from the
data, we have also computed the dipole solution splitting the sample
into field galaxies and groups/clusters. We find that for
$R\sim$~6000~\kms these sub-samples yield bulk velocities of
$\sim175$~\kms for groups/clusters and $\sim240$~\kms for field
galaxies, with errors of the order of $\sim 70$~\kms. These velocities
are somewhat higher ($\sim300$~\kms) when the objects are weighted by
their distance error. However, in this case the mean weighted depth is
small, for instance, $\sim2400$~\kms in the case of field
galaxies. The results obtained for field galaxies and groups/clusters
are, individually, in good agreement with the amplitude and direction
of the dipole obtained from TF surveys (Giovanelli \etal 1998a). We
conclude that on scales of $\sim6000$~\kms the sampling error is small
and comparable to the estimated random error of the bulk velocity
($\lsim$40~\kms). Adding this value in quadrature to that estimated
from the simulations we estimate the random error to be $\sim
60$~\kms. Another potential source of error in the bulk velocity are
systematic uncertainties in the distance arising from mismatches in
the velocity dispersion scale. Typically, the correction applied to
$\sigma$ for different runs is less than 0.020~dex with an uncertainty
of 0.009-0.018~dex, which in principle could lead to large errors in
the amplitude of the bulk flow. However, given the large number of
runs covering each region of sky and the fact that the observed
galaxies in each run were selected randomly, we estimate this
contribution to be at most $\sim 45$~\kms in each hemisphere. On the
other hand, the uncertainty in the offset between measurements of the
velocity dispersion from northern and southern observations is
estimated to be $\lsim$~0.006~dex, as determined from the a sample of
galaxies observed from both hemispheres. This uncertainty corresponds
to about 1.5\% in distance or to $\sim$ 50~\kms, which we take as an
an upper limit to the systematic error in the measured bulk velocity.

\section {Conclusions}
\label{summary}

Using a sample of 1274 early-type galaxies in 696 objects comprising
414 isolated galaxies and 282 groups/clusters, drawn from the recently
completed all-sky ENEAR redshift-distance survey, we have computed the
dipole component of the local velocity field to a depth of
$\sim6000$~\kms. Our main conclusion is that the streaming motion
amplitude of the ensemble of galaxies within the largest volume
considered is small $\sim$ 200~\kms. Similar small amplitudes are
obtained when the sample is split into isolated galaxies and
groups/clusters, indicating that sampling effects are relatively minor
on these scales. The amplitude and direction of the ENEAR dipole agree
well with the results obtained from similar analysis using the SFI
TF-survey. This is a remarkable result since these samples consider
galaxies of different morphological types sampling different regions
of space, were selected using different criteria, and the peculiar
velocities are derived using different distance relations.  Small bulk
velocities have also recently been obtained using new TF data
(Courteau \etal 2000) as well as other distance indicators (see Dekel
1999 for a review). If these results are confirmed, the peculiar
velocity field observed locally can easily be accounted for by the
currently popular cosmological models.

\acknowledgments{The authors would like to thank C. Rit\'e and
O. Chaves for their contribution over the years.  Special thanks to
E. Branchini. The bulk of the data used in this paper were obtained
from observations conducted at the European Southern Observatory (ESO)
and the MDM Observatory.}

%%%%%%%%%%%%%%%%%%%%%%%%% FIGURES %%%%%%%%%%%%%%%%%%%%%%%%%%%%%%%%%%%%%%%

\newpage

\clearpage

%%%%%%%%%%%%%%%%%%%%%%%%% TABLES %%%%%%%%%%%%%%%%%%%%%%%%%%%%%%%%

% NEW MB

\begin{table}
\begin{center}
\caption{Dipole Component of the Velocity Field\label{tab:bulk}}
\begin{tabular}{lccccccc}
\tableline \tableline
Sample    &  $N$ & $\vert {\bf  v}_b \vert$ & l & b & $\vert {\bf  v}_b \vert$  & l & b\\
    &  & (kms$^{-1}$) & (degree) & (degree)& (kms$^{-1}$) & (degree) & (degree)\\
\tableline
Objects   &  &  &   UNIFORM  &  &   &   WEIGHTED   &  \\
\tableline
$R<$2000  kms$^{-1}$   & 77 & 442 $\pm$   97   &   310 $\pm$  16 &
21 $\pm$ 10 &  446  $\pm$ 78  &  308 $\pm$  14 &   23 $\pm$  8 \\
$R<$ 4000  kms$^{-1}$   & 324 & 147  $\pm$ 62 &  306 $\pm$  18 &  9
$\pm$   14 &  350 $\pm$   47 &    301 $\pm$   10 & 16 $\pm$   7 \\  
$R<$  6000 kms$^{-1}$   &  656 &  220 $\pm$  42 & 304 $\pm$ 16 &  25
$\pm$    11& 298 $\pm$  38 &   299 $\pm$ 10 & 18 $\pm$  7 \\ 
\tableline
\end{tabular}
\end{center}
\end{table}

\void{
\begin{table}
\begin{center}
\caption{Dipole Component for Field Galaxies and Groups/Clusters}
\label{tab:bulk1}
\begin{tabular}{lccccccc}
\tableline \tableline
Sample  & $N$ &$\vert {\bf  v}_b \vert$ & $l$ & $b$ & $\vert {\bf  v}_b \vert$  & l & b\\
     &  & (kms$^{-1}$) & (degree) & (degree)& (kms$^{-1}$) & (degree) & (degree)\\
\tableline
Field    &  &  &   UNIFORM  &     & &  WEIGHTED    & \\
\tableline
$R<$  2000 kms$^{-1}$ & 40 &  565 $\pm$  149 &  317 $\pm$ 20 &  29 $\pm$  15 &
                    601 $\pm$  127 &  315 $\pm$ 18 &  34 $\pm$  13 \\
$R<$  4000 kms$^{-1}$ & 186&  164 $\pm$  93  &  315 $\pm$ 22 &  26 $\pm$  16 &
                    359 $\pm$  83  &  312 $\pm$ 15 &  32 $\pm$  11 \\
$R<$  6000 kms$^{-1}$ & 390 & 239 $\pm$  61  &  311 $\pm$ 23 &  31 $\pm$  13 &
                    343 $\pm$  67  &  310 $\pm$ 14 &  31 $\pm$  10  \\
\tableline
Groups/Clusters      &   & &   UNIFORM  &   &  &   WEIGHTED    & \\
\tableline
$ R<$ 2000  kms$^{-1}$ & 37 & 251 $\pm$  127 &  295 $\pm$ 35  & 21 $\pm$  13 & 
298 $\pm$     102    &   298  $\pm$     24    &   21  $\pm$     11    \\
$R <$ 4000  kms$^{-1}$ & 138 & 58 $\pm$  99  & 289  $\pm$ 31  & -11 $\pm$ 19 & 
406  $\pm$    66   &     287  $\pm$    12   &     3  $\pm$      7     \\
$R <$ 6000  kms$^{-1}$ & 266 & 175 $\pm$ 70 &  293  $\pm$ 31&   17 $\pm$ 17  & 
298   $\pm$   57     &   287   $\pm$   13     &   7    $\pm$  8     \\
\tableline
\end{tabular}
\end{center}
\end{table}
}

\begin{references}


\reference{} Borgani, S.,  Bernardi, M., da Costa, L. N., Wegner, G.,
Alonso, M. V., Willmer, C. N. A., Pellegrini, P. S. \& Maia, M. A. G.,
2000, ApJ, in press

\reference{} Branchini, E., Teodoro, L., Frenk, C. S., et al. 1999,
MNRAS, 308, 1


\reference{} Courteau, S., Faber, S. M., Dressler, A., \& Willick,
J. A. 1993, ApJ, 412, L51


\reference{} Courteau, S., Willick, J. A., Strauss, M. A., Schlegel,
D., \& Postman, M,  2000, ApJ, submitted  (astro-ph/0002420)

\reference{} da Costa, L. N., Freudling, W., Wegner, G., Giovanelli, R.,
Haynes, M. P., \& Salzer, J. J. 1996, ApJ, 468, L5

\reference{} da Costa, L. N., Bernardi, M., Alonso, M. V., Wegner,
G., Willmer, C. N. A., Pellegrini, P. S., Rit\'e, C., \& Maia,
M. A. G. 2000, AJ, in press (Paper~I)

\reference{} Dale, D. A., Giovanelli, R., Haynes, M. P., Campusano, L. E.,
 Hardy, E., \& Borgani, S. 1999, ApJL, 510, 11

\reference{} Dekel, A., Eldar, A., Kolatt, T., Yahil, A., Willick,
J. A., Faber, S. M., Courteau, S., \& Burstein, D.  1999, ApJ, 522, 1

\reference{} Dekel, A., in Cosmic Flows: Towards and Understanding of
Large-Scale Structure, eds. S. Courteau, M. A. Strauss, \&
J. A. Willick, ASP Conf. Ser., astro-ph/9911501

\reference{} Dressler, A., Lynden-Bell, D., Burstein, D., Davies,
R. L., Faber, S. M., Terlevich, R., \&  Wegner, G.  1987, ApJ, 313, 42

%\reference{} Eldar et al. 1999, in preparation
%\reference{} Geller, M. J., \& Huchra, J. P. 1983, ApJS, 52, 61 

\reference{} Giovanelli, R., Haynes, M. P., Herter, T.,  Vogt, N. P.
 da Costa, Freudling, W.,L. N., Salzer, J. J., \& Wegner, G, 1997, AJ, 113,53


\reference{} Giovanelli, R., Haynes, M. P., Freudling, W., da Costa,
L. N., Salzer, J. J., \& Wegner, G. 1998a, ApJ, 505, L91 

\reference{} Giovanelli, R., Haynes, M. P., Salzer, J. J., Wegner, G.,
da Costa, L. N., \& Freudling, W.  1998b, AJ, 116, 2632 

\reference{} Hudson, M. J., Smith, R. J., Lucey, J. R., Schlegel,
D. J., \& Davies, R. L. 1999, ApJ, 512, L79

\reference{} Lauer, T. R., \& Postman, M.  1994, ApJ, 425, 418

\reference{} Lynden-Bell, D., Faber, S. M., Burstein, D., Davies, R. L.,
Dressler, A., Terlevich, R., \& Wegner, G. 1988, ApJ, 326, 19


\reference{} Mathewson, D. S., Ford, V. L., \& Buchhorn, M. 1992, ApJS,
81, 413 


\reference{} Riess, A. G., Davis, M., Baker, J., Kirshner, R. P.
1997, ApJ, 488, 1

\reference{} Rubin, V. C., Roberts, M. S., Thonnard, N., Ford, W. K.,
 1976, AJ 81, 719

\reference{} Strauss, M. A., \& Willick, J. A. 1995, Physics Reports,
261, 271


\reference{} Wegner, G. \etal 1999, in Cosmic Flows: Towards and Understanding of
Large-Scale Structure, eds. S. Courteau, M. A. Strauss, \&
J. A. Willick, ASP Conf. Ser., astro-ph/9908354

\reference{} Willick, J. A. 1990, ApJ, 351, L5

\reference{} Willick, J. A., Courteau, S., Faber, S. M., Burstein, D.,
Dekel, A., \& Strauss, M. A. 1997, ApJS, 109, 333


\reference{} Willick, J. A. 1999, ApJ, 522, 647


\end{references}
\end{document}